# WORLDWIDE USE AND IMPACT OF THE NASA ASTROPHYSICS DATA SYSTEM DIGITAL LIBRARY

Michael J. Kurtz, Guenther Eichhorn, Alberto Accomazzi, Carolyn Grant, Markus Demleitner and Stephen S. Murray
Harvard-Smithsonian Center for Astrophysics, Cambridge, MA 02138

## ABSTRACT

The NASA Astrophysics Data System (ADS), along with astronomy's journals and data centers (a collaboration dubbed URANIA), has developed a distributed on-line digital library which has become the dominant means by which astronomers search, access and read their technical literature.
Digital libraries permit the easy accumulation of a new type of bibliometric measure, the number of electronic accesses ("reads") of individual articles.
By combining data from the text, citation, and reference databases with data from the ADS readership logs we have been able to create Second Order Bibliometric Operators, a customizable class of collaborative filters which permits substantially improved accuracy in literature queries.
Using the ADS usage logs along with membership statistics from the International Astronomical Union and data on the population and gross domestic product (GDP) we develop an accurate model for world-wide basic research where the number of scientists in a country is proportional to the GDP of that country, and the amount of basic research done by a country is proportional to the number of scientists in that country times that country's per capita GDP.
We introduce the concept of utility time to measure the impact of the ADS/URANIA and the electronic astronomical library on astronomical research. We find that in 2002 it amounted to the equivalent of 736 FTE researchers, or $250 Million, or the astronomical research done in France.

## 1. INTRODUCTION

The NASA Astrophysics Data System Abstract Service resides at the center of URANIA, the most sophisticated discipline centered bibliographic system ever developed (Boyce 1998), (Boyce 1996). The typical astronomer, on average, uses the ADS every day; ADS users read approximately five to ten times the number of papers which are read in all the (traditional) astronomy libraries of the world, combined.

A detailed technical and historical description of the ADS can be found in a set of four papers: Kurtz et al. (2000), Eichhorn et al. (2000), Accomazzi et al. (2000), and Grant et al. (2000).

The key fact which links the sections of this paper together is that the existence of digital libraries, like the ADS, means that detailed information on the readership of journal articles, on a per article and per reader basis, can now be routinely obtained. This is basically a new bibliometric measure of importance equal to citations. A detailled examination of the bibliometric properties of the readership information, both in comparison to and in conjunction with citation information can be found in our companion article Kurtz, et al. (2004)(hereafter Paper2).

In section 2 we briefly describe the system, and note some of its more important or sophisticated features. In section 3 we breifly describe a power set of collaborative filters (e.g. Goldberg, et al. (1992)), which make use of the new information to achieve improved information retrieval, we call these second order operators,

In section 3 we describe the growth and current use of the system, and in section 5 we use ADS usage to show the relationship a country's wealth and population has with the number and effectiveness of its scientists.

Finally in section 6 we discuss how the total impact of the ADS and the electronic astronomy library can be evaluated; and we conclude in section 7 by discussing some of the future consequences of the new technologies and data used here.

## 2. DESCRIPTIVE BACKGROUND

The ADS Abstract Service was first demonstrated in 1992 (Kurtz et al. 1993) and was put on-line for general use in April 1993. In early 1994 the service was moved onto the World Wide Web from its original NASA developed network home (Murray et al. 1992). In this section we will briefly describe some of the more important milestones and features in the ADS development; Kurtz et al. (2000) contains a detailed description of the early history of the ADS Abstract Service.

The ADS from the beginning has been designed to be used primarily by end-user astronomers, as opposed to librarians. This has led us to create a system which emphasizes recall over precision, assuming that the expert user can easily separate the desired from the undesired. The behavior of the system can be easily modified from its default, however, to match almost any desired searching style.

From the beginning our standard search involved partial match techniques, natural language queries with an extensive discipline specific synonym list, relevance feedback, and the combination of evidence (Belkin, et al. 1995) from simultaneous queries, such as text from the abstract combined with text from the title, combined with an author's name.

Within a few months of the initial public release, in 1993, we were able to include the ability to query for papers containing information about a particular astronomical object (i.e. a particular star or galaxy) by perform-ing simultaneous joint queries with the SIMBAD data base (Wenger et al. 2000), operated by the CDS (Centre des donnees astronomique de Strasbourg) in France. We believe this is the first time the internet was used to permit routine joint transatlantic queries of this sort. This collaboration continues, and forms the core of URANIA (Kurtz et al. 2000).

Also, soon after moving to a web based system, we were able to make direct hyperlinks between the bibliographic entries and on-line versions of data tables from those articles from the journal *Astronomy & Astrophysics* kept by the CDS in Strasbourg (Ochsenbein and Lequeux 1995). This capacity has grown so that now essentially every data table from every article in every major journal of astronomy is available via these means.

Towards the end of 1994, the ADS began scanning (creating bi-tonal bitmapped images of) recent issues of *The Astrophysical Journal (Letters)*. With the collaboration of virtually every astronomy journal we now have electronic copies of nearly every astronomy research paper published in a journal during the last two hundred years, and with the collaboration of the Center for Astrophysics' Wolbach Library and the Harvard University Library (Corbin and Coletti 1995) we expect in the near future to have a nearly complete collection of the astronomy research literature going back two centuries. This collection has been called "Earth's Largest Free Full-Text Science Archive" (Highwire 2001).

In July 1995, *The Astrophysical Journal (Letters)* published its first issue in HTML electronic format (Boyce 1995), (Dalterio et al. 1995); it was the first major scientific journal to be fully on-line. The reference sections of that first electronic issue had hyperlinks from the references to the corresponding entries in the ADS. This practice of linking from the electronic article references to the relevant sections of a bibliographic service has now become standard for electronic journals; now all astronomy journals, and several others, link to the ADS.

Also in 1995 the ADS made its first links between a bibliographic reference and the data which were used to do the research described in the article (Plante et al. 1996) . This practice has grown tremendously; more than 175,000 such links are now in the ADS, and the growth is becoming more rapid as part of the emerging Astronomical Virtual Observatory (Brunner et al. 2001), (Kurtz and Eichhorn 2001).

In 1996, using data purchased from the Institute for Scientific Information (ISI) with funds provided by the American Astronomical Society (AAS) as part of the electronic *Astrophysical Journal* project, we provided links from the main article references to lists of the papers they referenced, and to the papers which cite them (Kurtz et al. 1996). With the cooperation of all the main journals of astronomy we have since greatly expanded this capability, by purchasing more data from ISI, by performing optical character recognition on the reference sections of scanned articles in our database, and by taking directly the references from electronic articles provided to us by the publishers (Demleitner et al. 1999). Currently there are more than 12,000,000 citing-cited article pairs in our database, and the number continues to grow.

In 1996, we began maintaining mirror sites around the world (Eichhorn et al. 1996); the first was hosted by the CDS in Strasbourg. Currently we have twelve mirrors in twelve countries on four continents.

In 2000, we added the ability to see articles which were the most read by people who read another article (i.e. people who read this article also read: . . ., hereafter also-read). Because ADS users read a substantial fraction of all astronomy articles (see sections 4 and 6) this feature provides an accurate sampling of the entire discipline.

## 3. SECOND ORDER OPERATORS

In 1997, we began permitting some second order queries (Kurtz 1992) by allowing the result of a query to be the collated reference or citation lists of the articles returned by the query, rather than the articles themselves (Kurtz et al. 2000) . In 2001 we added the also-read lists to this capability, and created a new interface, both to make it more intuitive, and to improve its functionality. The resulting set of second order operators provides an extremely powerful basis for creating specialized collaborative filters (e.g. Goldberg, et al. (1992)). We believe many of these capabilities are unique to the ADS.

A full description of the second order operators will be published elsewhere (Kurtz et al. 2002); here we will give one example of their use, in order to demonstrate their power.

We will begin by assuming that we are interested in the current state of knowledge in subfield X. We will further assume that we know enough about subfield X (by knowing an author's name, perhaps) to be able to find the abstract of at least one paper about subfield X.

We retrieve the abstract(s) using the ADS and use the "find similar abstracts" feature, modifying the query to return only articles published within the past year, we then retrieve a list of the recent papers with the most similar abstracts. This is a list of the most recent papers in subfield X based on similarity of text. We can now use the second order operators.

The people who have recently read the most recent papers on subfield X are people who are interested in subfield X; the papers they read the most will be the currently most popular papers on subfield X. Clicking on the "get also-read lists" feature retrieves exactly those most popular articles on subfield X.

These papers are the most popular for a reason; presumedly they are informative papers concerning the current state of affairs of subfield X. The papers most referenced by these papers would be those currently most useful to subfield X; clicking on the "get reference lists" feature gets this list of most useful (most cited by papers in subfield X) articles.

Papers which cite a large number of these most useful articles are papers with extensive discussions of topics of interest to subfield X; they are the most instructive articles about subfield X. Clicking on the "get citation lists" feature gets this list of most instructive articles. Figure 1 shows this list after following the above procedure starting with the abstract from the paper by Riess et al. (1998) "Observational Evidence from Supernovae for an Accelerating Universe and a Cosmological Constant." The score refers to the number of references in the previous list (of 500) which are also cited by these articles. For example, the top article in this list cited 96 of the papers from the previous list of most referenced papers; for three of the five papers shown more than half of all the papers in their reference lists were also in the previous list. All of the articles on the top of the list are review articles on exactly the original subject; the high quality of this list tends to validate each stage of the process.

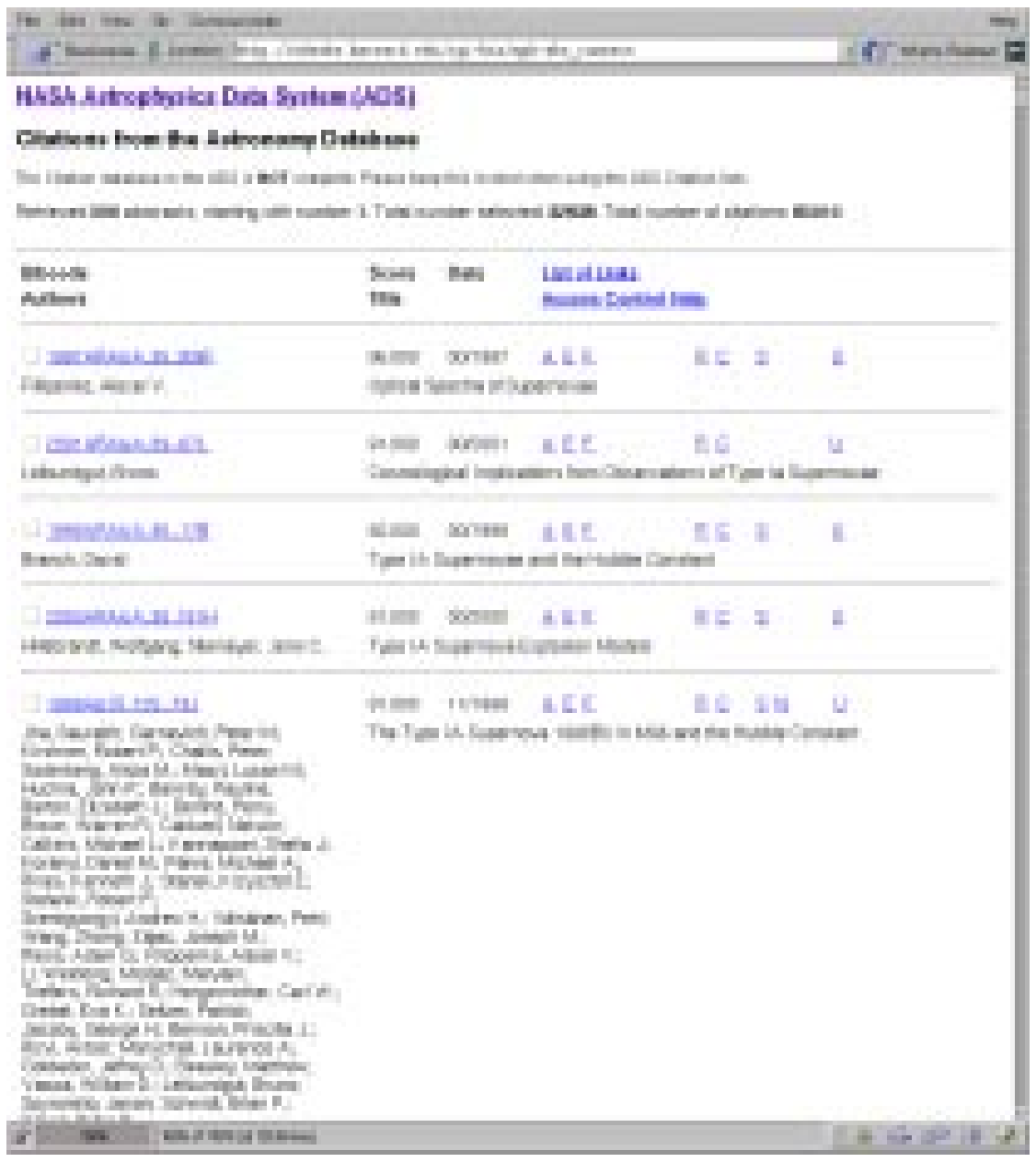

Fig. 1.— The "most instructive" papers list obtained by following the procedure described in the text, beginning with the abstract from Riess et al. (1998).

Beginning with a minimum of knowledge (the ability to find a single article on the desired subject) we are quickly able to obtain lists of the most current, most popular, most useful, and most instructive articles about the desired subject. Given the ability to write a good query, or to edit the lists to include only the most relevant articles, one can obtain in-depth results for very narrow subject matters. Of course very many different ways of using these second order operators are possible.

While we believe our use of these second order operators is unique, there are some similar features in other systems. The "customers who bought this book also bought" feature of Amazon.com (Amazon 2002) and other on-line retailers is similar to using the "get also-read" feature on a single article. The "get related records" feature of ISI's Web of Science (ISI 2002) is the same as running the "get citation lists" feature on the reference list from a single article.

## 4. THE GROWTH OF THE ADS

The ADS went on-line in April 1993 using a wide area networking system created for NASA; in its ten months of existence usage approximately doubled. In February 1994, the ADS converted to the World Wide Web (Grant et al. 1994); usage went up by a factor of four within five weeks. Since then use of the ADS has doubled about every two years. There are a number of different ways to measure ADS use, figure 2 shows the number of unique users per month. The dotted line shows a two-year doubling.

We caution that the number of unique users may be misleading. We track usage by the number of unique "cookies"[1] which access the ADS, and by the number of unique IP[2] addresses. This leads to a number of unique users which may exceed the actual number of different people using the system; a person who logs in from work and from home would count twice, but any number of people using a library computer would count as one; also a person who used more than one of our dozen mirror sites would be counted more than once. The monthly usage for October 2001 was 51,960 users.

This number may be compared to 8,692 members of the International Astronomical Union (IAU) (IAU 2001), the largest organization of professional astronomers. There were 10,677 different authors listed in the papers published in 2001 by the seven major astronomy journals, 5576 of these were listed as (co-)author on more than one paper. Independent of how many actual users the ADS has, we believe nearly every working astronomer uses the ADS very regularly.

Table 1 compares the number of unique users who read ten or more articles during August 2001 using the ADS

Table 1. Use by Selected Institutions, August 2001

| Institution | heavy users[a] | AAS Members[b] | unique reads[c] |
|---|---|---|---|
| Harvard-Smithsonian | 236 | 250 | 8085 |
| Space Telescope | 126 | 163 | 5271 |
| CalTech | 173 | 196 | 5589 |
| Cornell | 33 | 52 | 899 |
| European Southern Obs. | 385 | 18 | 11274 |
| Total worldwide | 10532 | 6429 | 350404 |

[a]Unique users who have read ten or more articles during the month
[b]AAS (2001)
[c]Number of unique user-paper pairs

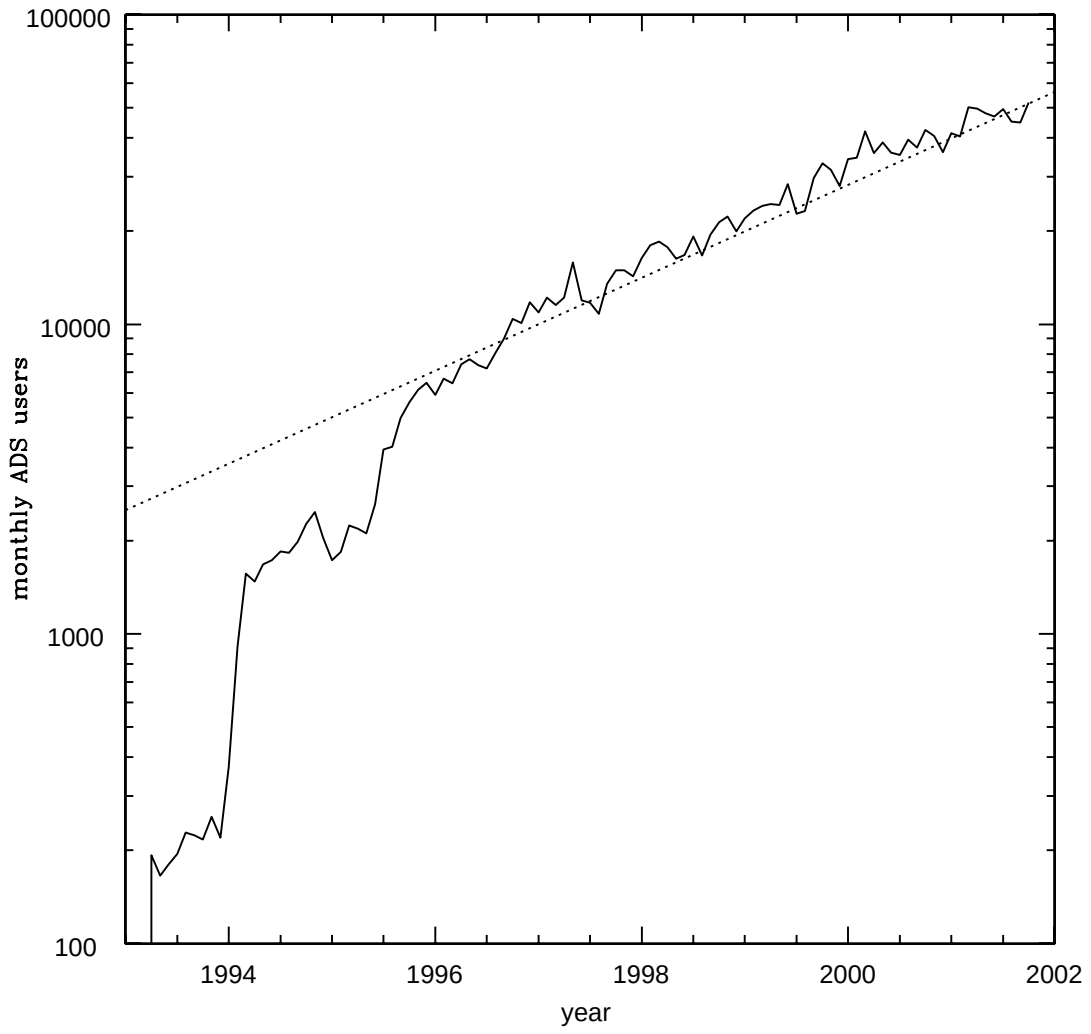

Fig. 2.— The number of unique users of the ADS each month. The dotted line represents a 2 year doubling time.

[1] A cookie is a unique identifier which WWW providers (in this case ADS) assign to each user, and store on the user's computer using the browser.

[2] Each Machine on the internet has a unique IP (Internet Protocol) address.

to the number of members of the American Astronomical Society for some selected astronomical institutions. The number of AAS members is similar to the number of heavy ADS users for the US organizations; more than twice as many users are present at each institution, but they use the ADS less frequently, the heavy users account for about 85% of the total use of the system.

The number of unique reads represents the number of unique user-article pairs in a month, thus if a single user reads the same article twice, it only counts once. This is the most conservative estimate of number of articles read in the system. For heavy users the median number of articles read per month is 22, with very little change among various organizations or subgroups. This is the basis for our claim that the typical working astronomer, on average, uses the ADS every working day.

As the numbers for the European Southern Observatory, which is located near Munich, demonstrate, the use of the ADS is not limited to organizations within the United States. Three fifths of all ADS use comes from outside the United States; this will be explored in the next section.

Tenopir, et al. (2003) independently discuss the degree to which the ADS is used by astronomers.

## 5. ADS AS INDICATOR OF THE WORLDWIDE RESEARCH EFFORT

Astronomy is a worldwide enterprise, and, as probably the least practically useful of all sciences, is perhaps the best indicator of how pure research is done internationally. (There is no such thing as "Applied Astronomy"). As the ADS use is overwhelmingly by astronomy researchers, a country by country comparison of ADS use can yield insights into the factors which influence the world's pure research.

### 5.1. *ADS use and GDP*

Certainly one factor which influences the research done in a country is the wealth of that country. Figure 3 shows the ADS use per capita vs. the GDP (Gross Domestic Product) per capita. There is clearly a trend. The center line shows the relation where ADS use per capita is proportional to GDP per capita squared; the lines to each side are separated by a factor of $\sqrt{2}$ in per capita GDP.

Taking ADS use as a proxy for basic research we obtain directly a simple relation for the amount of basic research done in a country:

$$BasicResearch \propto (GDP)^2/Population \qquad (1)$$

The ADS use is obtained by counting the number of entries in the query log (this is approximately twice the true number of requests to the system) made during the year 2000, and the GDP is by the purchase power parity method (CIA 2001). Most queries can be automatically resolved to the internet host address, but about 20% cannot; the IP addresses which do not resolve are systematically more frequently found in the less industrialized countries. We have resolved by hand (to the country level) all class B and class C IP subnets with more than 1,000 queries during the year 2000, this procedure only significantly changed the data for two countries, Egypt and Morocco. By this process we can be certain that we have not missed any large group of astronomers (1,000 queries or about two true requests per work day, twice the median use of the typical user(section 4), is a typical amount for a single very active user); we have not resolved about 10% of the queries, and can be certain that the results for some countries with low usage are systematically underestimated.

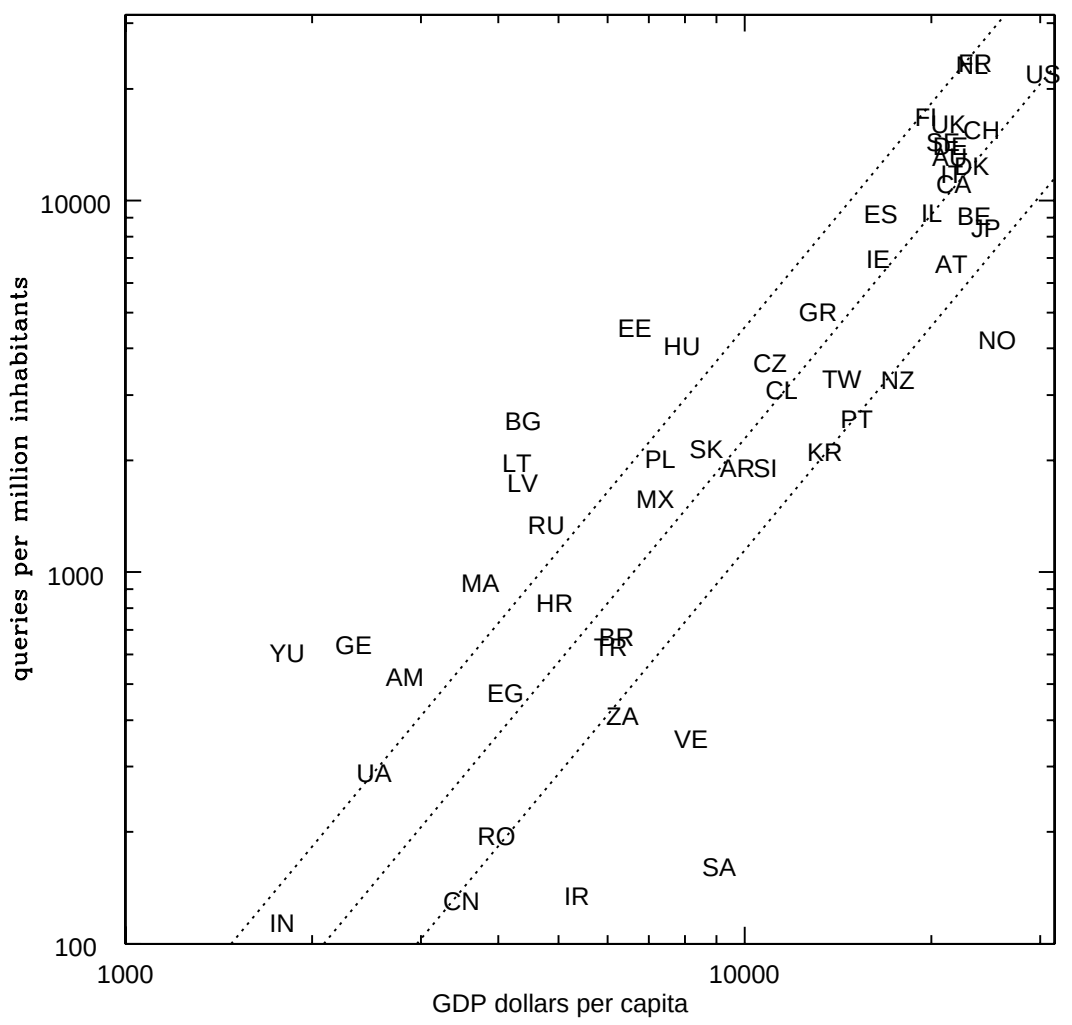

FIG. 3.— The per capita GDP of a country vs. the ADS use per million inhabitants in that country. The symbols are the ISO (International Standards Organization) symbols for each country.

Comparing the GDP figures from the CIA (CIA 2001) with those from the Economist Magazine (Economist 2001) shows that these can differ by a factor of two for some developing countries, and by as much as 30% even in industrial countries.

Figure 3 contains only countries where the number of ADS queries is equal or greater than 1,737, which is the average number of queries per member of the IAU, and is about equal to the use of two active users. There are several things which can be seen in the plot. First, notice that the United States is not the largest per capita user of the ADS; France (FR) and the Netherlands (NL) are; their symbols obscure each other on the plot. Also notice that while the difference in per capita income between the richest and poorest countries is a factor of ten to fifteen, the difference in per capita ADS use between these countries is close to three hundred.

Next notice that the countries to the left of the dotted lines, those which perform more research per capita than their GDP per capita squared would suggest, are, with one exception (Morocco, MA) all eastern European countries which have undergone substantial economic and political change in recent years. It is possible that in the face of such large historical changes GDP is not a fully accurate measure of the total wealth of a country, which would also include existing infrastructure.

Also notice that the four countries to the right of the dotted lines, those which perform less research than expected, are all oil producing states. It is possible that the (historically) recent increased GDP due to oil revenues in these countries is not a true measure of their actual long term wealth.

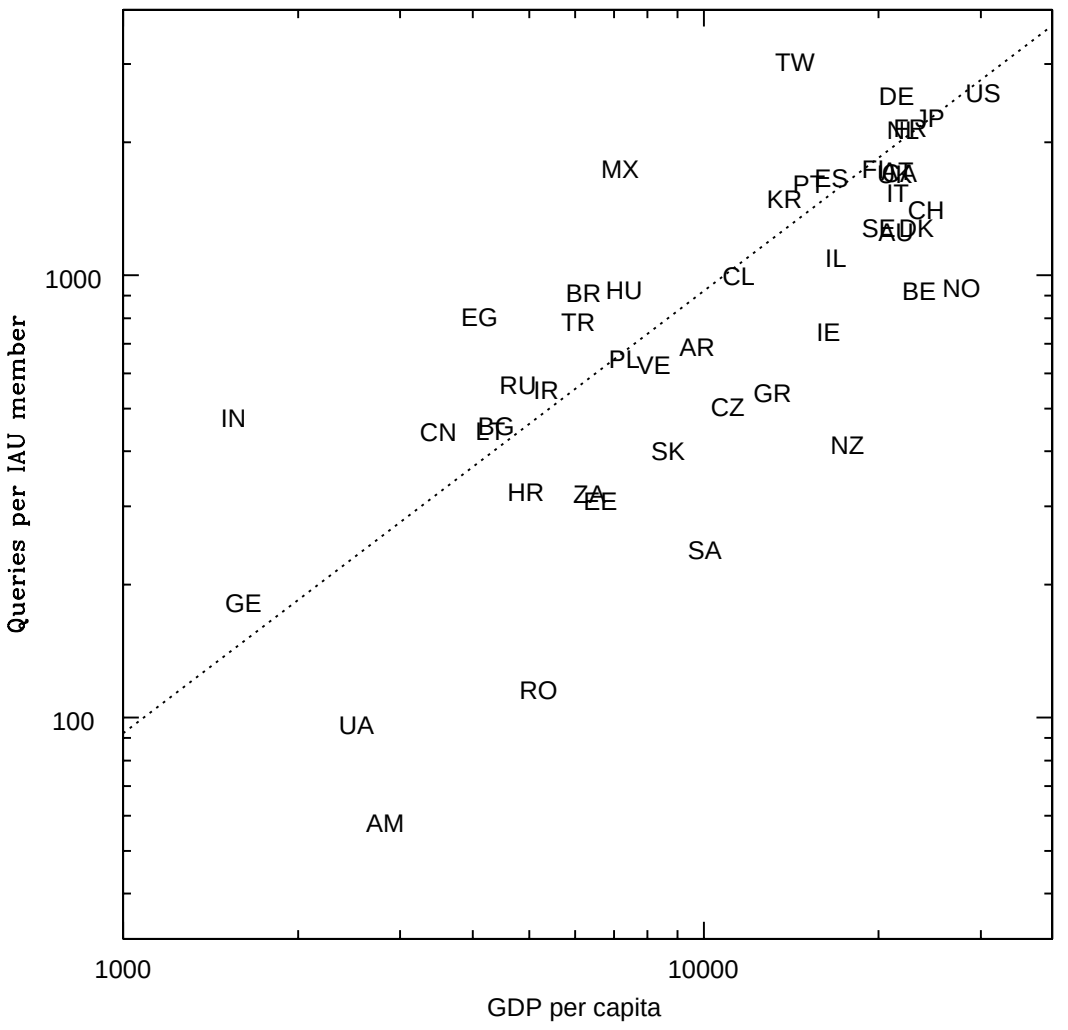

FIG. 4.— The per capita GDP of a country vs. the ADS use per the number of IAU members in that country. The symbols are the ISO symbols for each country.

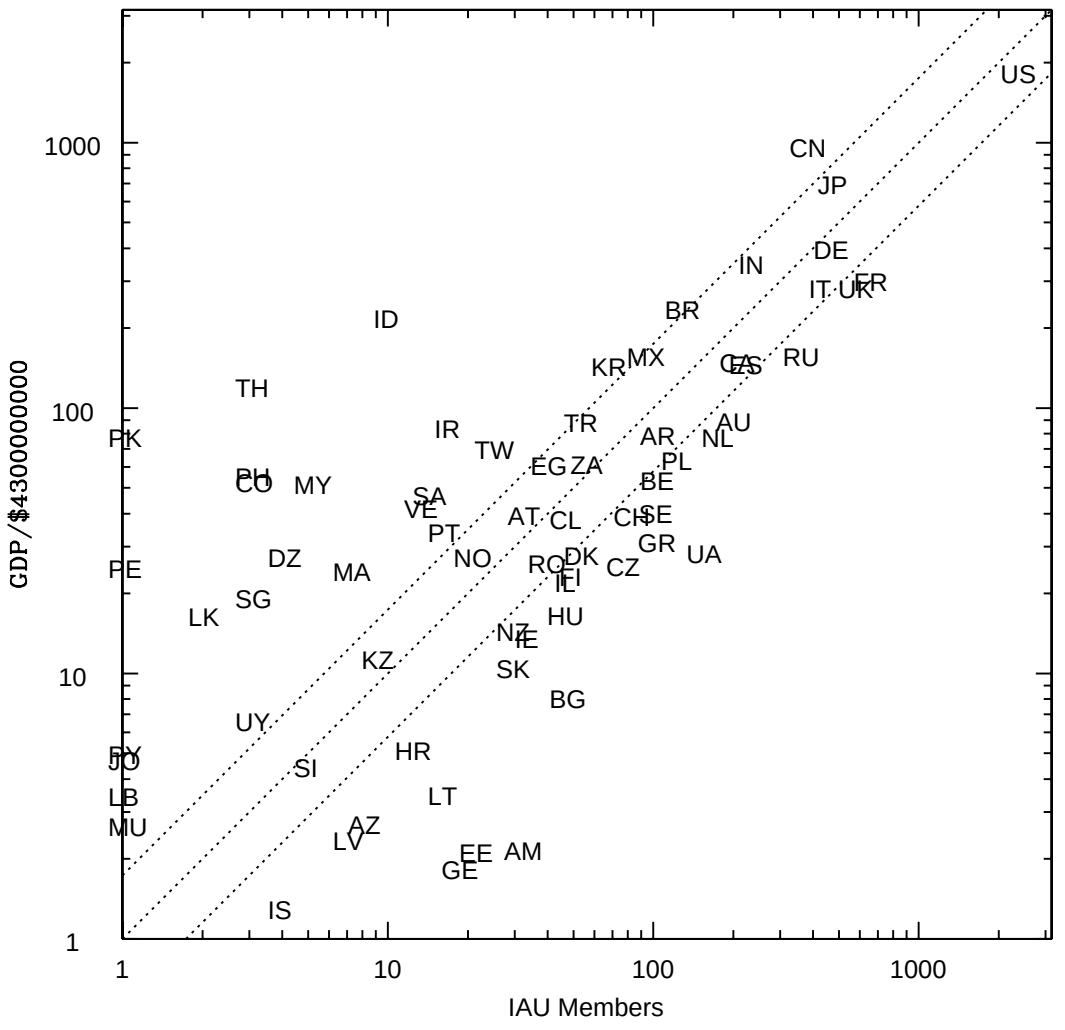

FIG. 5.— The GDP of a country in units of $4,300,000,000 vs. the number of IAU members in that country. The symbols are the ISO symbols for each country.

### 5.2. *ADS use and IAU membership*

Another factor which affects the amount of research done in a country is the number of researchers in that country. Again using astronomy as an indicator of the pure research effort we will take the membership in the IAU on a per country basis (IAU 2001) and compare that to the amount of ADS use in each country.

One might naively assume that on average the per astronomer use in each country would be constant; the average astronomer would do the average amount of research independent of location. However, figure 4 shows for each country the number of ADS queries per IAU member as a function of GDP per capita. In fact there is a clear trend, the dotted line shows the relation where ADS use per IAU member is proportional to GDP per capita, hence the productivity of individual astronomers is not independent of location. The dotted line is not a fit, and the scatter in this relation is too large to derive a parameterization. The dotted line is perhaps the simplest relation which is consistent with the data.

### 5.3. *IAU membership vs. GDP*

Combining the relations shown in figure 3 (ADS use per capita is proportional to GDP per capita squared) and figure 4 (ADS use per IAU member is proportional to GDP per capita) one obtains the relation that IAU membership is proportional to GDP. This relation can also be obtained in other ways, and is obviously independent of the ADS. Figure 5 shows the relation; the central dotted line represents the relation one astronomer per $4,300,000,000 which is the total world GDP divided by the number of IAU members.

The trend in these data is clear. Nearly all countries on the plot follow the relation, the principal exceptions being Indonesia, Thailand, and Pakistan (ID, TH, and PK). The countries, however, do not seem to form a single linear relation. The two flanking lines, which are separated from each other by a factor of three, seem to better describe the distribution than the single average. We suggest that this bifurcation is real, and represents the cultural difference between western nations and eastern nations with regard to support of basic research. Additionally some countries perform essentially no astronomy; they are primarily from southeast Asia (excluding India) and subsaharan Africa.

If one divides the data by the mean line down the center, one finds that 31 European countries are below the line and three above. One also finds that seventeen Asian countries are above the line, and none are below. We thus propose:

$$Scientists = \frac{c}{k} GDP \tag{2}$$

where for astronomy

$$c = \begin{cases} \sqrt{3} & \text{if european culture;} \\ \frac{1}{\sqrt{3}} & \text{if asian culture;} \\ 0 & \text{otherwise.} \end{cases}$$

and

$$k = \$4,300,000,000$$

*Scientists* is the number of scientists in a country, GDP is the GDP of that country, k is the mean GDP per scientist worldwide, and c is a cultural factor which differs among countries.

### 5.4. *A model for worldwide basic research*

Rearranging the relation shown in figure 4 we obtain the following empirical model for the research done in a country:

$$BasicResearch \propto Scientists \frac{GDP}{Pop} \tag{3}$$

The amount of basic research done in a country is proportional to the number of scientists times the per capita GDP.

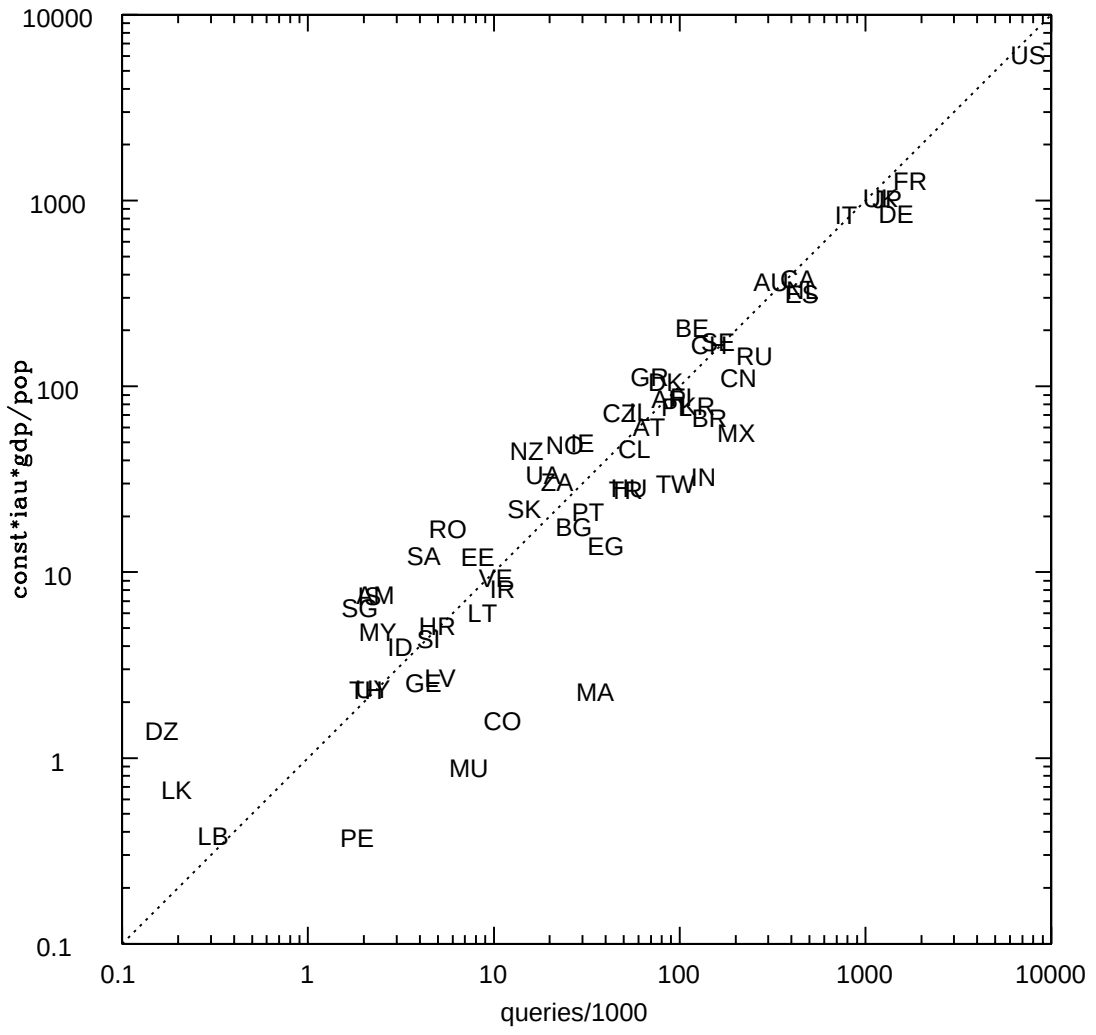

FIG. 6.— The number of astronomers times the per capita GDP vs. ADS use for each country. The symbols are the ISO symbols for each country

Figure 6 shows this model; the dotted line represents the same mean proportionality constant as in the line in figure 4. The model in equation 3 provides a reasonably accurate prediction of the research output of a country for countries differing in research output over a factor of ten thousand.

### 5.5. *Discussion*

Combining equations 2 and 3 we get a simple picture of the major factors affecting research worldwide. The number of scientists in a country is just proportional to the GDP of that country, except that Asian and European countries have a different proportionality constant, by a factor of three (and some countries simply do no astronomy); and the amount of research done in a country is proportional to the number of scientists in that country, times an efficiency factor which is just the GDP per capita.

This model for the research of nations is essentially a quantification of the Matthew Effect for Countries (Bonitz 1997), an extension of the Matthew Effect (Merton 1968) which essentially states that the system of rewards in science is that the rich get richer and the poor stay poor.

According to this model two countries with the same GDP (for example India and France) will have the same number of scientists (ignoring the cultural factor). The amount of research these scientists perform, however, differs by the ratio of the per capita GDP for the two countries, more than a factor of ten in the case of India and France.

Other studies have jointly looked at bibliometric and economic measures; for example May (1997) has shown the research expenditure (in £1,000,000 per citation) for a number of countries. Table 2 shows the the percentage of total reads, as well as citations and number of papers (National Science Board 2002) for some selected countries.

The percentage of astronomy reads per country actually predicts the average of the percentage of citations and scientific papers (which are for all fields) better than the citations and publications predict each other. For half of the 47 countries which produce more than a five thousandth of the world's total citations the percentage of astronomy reads is within ±19% of the average percentage of citations and number of publications over all fields. For half of these countries the percentage of number of publications is within ±36% of the percentage of citations.

TABLE 2. COMPARISON OF BIBLIOMETRIC MEASURES SELECTED COUNTRIES

| Country | Reads[a] | Cites[b] | Papers[c] |
|---|---|---|---|
| Argentina | 0.5 | 0.2 | 0.4 |
| Brazil | 0.8 | 0.4 | 1.0 |
| Canada | 2.3 | 4.0 | 3.7 |
| China | 1.2 | 0.6 | 2.2 |
| Germany | 7.9 | 7.0 | 7.1 |
| India | 0.8 | 0.6 | 1.7 |
| Japan | 7.3 | 7.1 | 9.0 |
| Mexico | 1.1 | 0.2 | 0.4 |
| Poland | 0.5 | 0.4 | 0.8 |
| Russia | 1.3 | 0.9 | 3.0 |
| United Kingdom | 6.5 | 8.0 | 7.5 |
| United States | 40.5 | 52.1 | 30.9 |

[a]Percent of worldwide total (2001)
[b]Percent of worldwide total (1999),National Science Board (2002)
[c]Percent of worldwide total (1999),National Science Board (2002)

## 6. THE IMPACT OF THE ADS AND THE ELECTRONIC ASTRONOMY LIBRARY

The ability to access information rapidly and easily is transforming modern civilization. Many have compared the recent advances in electronic communications with the invention of the printing press in terms of historical importance. The widespread use of the ADS within the URANIA collaboration (Boyce 1998) has clearly had a profound effect on how astronomical research is performed; in this section we attempt to quantify the magnitude of that effect.

Expanding on the discussion in Kurtz et al. (2000), we define the concept of utility time as a measure of how a tool improves efficiency. Utility time is the time which would have to have been spent to achieve the same result, were the tool not present. For a research tool such as the ADS this is the amount of research (or researcher's) time gained by using the ADS (together with the URANIA collaboration) compared with not using it.

The principle "product" of the ADS is access to articles in research journals. It is difficult to determine the research time gained by reading an article in a research journal. It can range from zero (the article was worthless, so the time was wasted) to years (the proposed study has already been done). We propose, on the average, however, that the time gained by reading the technical literature must be positive, and at least equal to the time spent getting and reading it; otherwise rational scientists would not waste their time reading the literature.

The ADS combined with the electronic journals essentially eliminate the time required to get an article. We estimate this by timing the first author (MJK) as he writes a reference on a small piece of paper, walks two flights downstairs to the library, searches the stacks for the proper volume (it was an *Astrophysical Journal* reference), takes the volume to the photocopy machine and photocopies the article, then finally walks back upstairs to his office. 15 minutes was the elapsed time, and we take that as the utility measure in research time gained by using the ADS to obtain a full text article from the technical literature.

Obviously much of the credit for the ease of access to the electronic literature goes to the electronic journals themselves. In astronomy the electronic journals and the ADS have grown up together in a symbiotic relationship. Here we make no attempt to split credit for what have been joint developments. Credit for the utility of the contents of the journals is a separate issue, and clearly belongs to the journals. The collaboration of the journals with the ADS, discussed here, represents an important fraction of the URANIA collaboration; the collaborations surrounding the CDS in Strasbourg (Genova et al. 2000) represent another large portion, and would have to be added in to ascertain the impact of the full URANIA project. The expansion of URANIA to include observational data in observatory archives promises to be even more valuable. These projects now use the name "virtual observatory."

We also assign one third of the utility value for whole text retrieval (5 minutes) to the retrieval of an abstract, reference list, or citation list. In practice this is probably an underestimate.

Finally we arbitrarily assign a one minute benefit for each simple query made to the system. Even a very common query, such as 'show me recent papers by author X' would take longer than a minute to get by any pre-ADS means, while complex queries, such as the joint ADS–SIMBAD query 'show me papers about the metallicity of the globular clusters in the galaxy M87' would have taken from hours to weeks previously.

The result for the research time gained using the ADS/URANIA during the year 2002 is shown in table 3. The table shows the types of data available via the ADS for articles, how many times each type was accessed in 2002, the utility value assigned to each type, and the total research time gained, measured in 2,000 hour full time equivalent researcher years. For a detailed description of all the various data types see Eichhorn et al. (2000). The bottom line is that the ADS/URANIA improved astronomical research during 2002 by the equivalent of 736 full time researchers.

To put 736 FTE researchers into perspective: The Harvard-Smithsonian Center for Astrophysics, the largest astronomical research facility in the United States, has about 300 research scientists, and has an annual budget of about \$100 Million. If the ADS/URANIA provides $2\frac{1}{2}$ times the research scientists as the CfA its annual value to astronomy research can be calculated as $2\frac{1}{2}$ times the CfA budget, or \$250 Million. The ADS yearly budget is but a tiny fraction of that sum. $2\frac{1}{2}$ times the CfA is also approximately the astronomical research output of the country of France.

From another perspective we can estimate that there are ∼12,000 astronomers in the world. 736 FTE astronomers would then represent ∼6% of the worlds astronomers. Woltjer (1998) estimates that the yearly budget for astronomy research, worldwide, is between \$4000 and \$5000 Million. 6% of that would be between \$240 and \$300 Million, essentially the same as in the preceding paragraph.

Table 3. ADS Use 2002

| Code | Function[a] | Number | Time[b] | Research Gain[c] |
|---|---|---|---|---|
| A | Abstract | 4,171,704 | 5 | 174 |
| C | Citations | 676,305 | 5 | 28 |
| D | Data | 61,678 | | |
| E | HTML Article | 864,019 | 15 | 108 |
| F | PDF Article | 1,872,035 | 15 | 234 |
| G | GIF Article | 677,821 | 15 | 85 |
| I | Author Comments | 748 | | |
| L | Library Entry | 2,682 | | |
| M | Document Delivery | 9,322 | | |
| N | NED Entry | 33,963 | | |
| O | Associated Articles | 10,373 | | |
| P | PDS Entry | 76 | | |
| R | Reference List | 152,188 | 5 | 6 |
| S | SIMBAD Entry | 97,691 | | |
| T | Table of Contents | 47,373 | | |
| U | Also Read | 150,891 | | |
| | Simple Queries | 12,168,336 | 1 | 101 |
| | | | | 736 |

[a]See Eichhorn et al. (2000) for a full description
[b]Minutes of research time gained per query
[c]In 2000 hour FTE Research Years

Another way to view the impact of the ADS and the electronic journals on astronomy research is to compare the total readership of astronomy's technical literature now with what it was previously. Kurtz et al. (2000) estimated the total number of astronomy research articles read in traditional libraries at about 600,000 per year. If one ignores articles read in the browse mode upon arrival of a new journal (the N mode of section 2.1 of Paper2) the number of articles read by individuals with personal subscriptions must be less than the total read in libraries. We will estimate that no more than 1.2 million papers per year are read by astronomers using non-electronic means.

Table 3 shows that 3.4 million full text articles have been accessed using the ADS during 2002, nearly three times the total number of non-electronic reads. In addition 4.2 million abstracts were accessed and 0.8 million citation and reference lists; this represents an entirely new mode of rapid information gathering on the part of researchers. The ADS and the electronic astronomical library have succeeded in increasing the total readership and use of the technical astronomical literature by a factor of at least three within a very short time; this represents a massive change in the research behavior of working astronomers today.

## 7. CONCLUSIONS

As the first digital science, astronomy has long been a leader in the information technologies which are transforming our civilization. With respect to the astronomical technical literature the close collaboration of the data centers, the journals, and the ADS has created URANIA, a fully integrated system of enormous power; within just a few years it has become universally used.

At the heart of this system, the ADS presides over a huge database, with many of the data-sets being brought together for the first time. Simple queries, such as 'show me the most cited papers containing the phrase "redshift survey"' were not possible before the ADS merged the text and citation databases in 1997.

We have demonstrated, in section 3, that very sophisticated and powerful retrieval algorithms can be created by merging text, citation, and reference databases with a database of articles and their readers. Surely numerous new techniques will be developed in the coming years taking advantage of the tremendous possibilities inherent in these data. Recently Bollen, et al. (2003) have used a similar comparative technique to identify current research trends.

In section 5 we showed that the correlation of readership information with place can permit large scale sociometric studies of the research efforts of countries, as well as smaller entities. The ease and rapidity with which this information can be gathered will not only permit similar studies in other fields, but will also permit changes in relative activity to be studied in nearly real-time.

The number of research articles published in astronomy has increased by a factor of four in the time since astronomers now on the verge of retirement entered the field. For years this explosion of information has proved a problem, as researchers were able to read a smaller and smaller fraction of the published literature. With the advent of the new information technologies inherent in the electronic astronomy library, and centered about the ADS, researchers are now able to find and read a substantially larger fraction of what is published; essentially they are now able to read the same fraction as forty years ago. These technologies have changed, and continue to change, the practice of scientific research in a fundamental way.

## 8. ACKNOWLEDGMENTS

It is a pleasure to thank R. Scheiber-Kurtz for extensive discussions, and for detailed reading and criticism of the text; to thank Edwin Henneken for reading the text; and to thank M.J. Geller for discussions.

The ADS is supported by NASA under Grant NCC5-189.